\documentclass[longauth]{aa} 

\usepackage{graphicx}
\usepackage{txfonts}
\begin{document}

   \title{The dust environment of comet 67P/Churyumov-Gerasimenko 
from Rosetta OSIRIS and 
VLT observations in the 4.5 to 2.9 au heliocentric distance range inbound}


\author{
F. Moreno\inst{\ref{inst1}}
\and C. Snodgrass\inst{\ref{inst50}}
\and O. Hainaut\inst{\ref{inst51}}
\and C. Tubiana\inst{\ref{inst3}}
\and H. Sierks\inst{\ref{inst3}}
\and C. Barbieri\inst{\ref{inst4},\ref{inst5}}
\and P. L. Lamy\inst{\ref{inst6}}
\and R. Rodrigo\inst{\ref{inst7},\ref{inst8}}
\and D. Koschny\inst{\ref{inst9}}
\and H. Rickman\inst{\ref{inst10},\ref{inst16}}
\and H. U. Keller\inst{\ref{inst11}}
\and J. Agarwal\inst{\ref{inst3}} 
\and M. F. A'Hearn\inst{\ref{inst12}}
\and M. A. Barucci\inst{\ref{inst13}}
\and J.--L. Bertaux\inst{\ref{inst14}}
\and I. Bertini\inst{\ref{inst5}}
\and S. Besse\inst{\ref{inst9}}
\and D. Bodewits\inst{\ref{inst12}}
\and G. Cremonese\inst{\ref{inst15}}
\and V. Da Deppo\inst{\ref{inst17}}
\and B. Davidsson\inst{\ref{inst16}}
\and S. Debei\inst{\ref{inst18}}
\and M. De Cecco\inst{\ref{inst19}}
\and F. Ferri\inst{\ref{inst5}}
\and S. Fornasier\inst{\ref{inst13}} 
\and M. Fulle\inst{\ref{inst21}}
\and O. Groussin\inst{\ref{inst6}} 
\and P.J. Guti\'errez\inst{\ref{inst1}}
\and P. Guti\'errez--Marques\inst{\ref{inst3}} 
\and C. G\"uttler\inst{\ref{inst3}} 
\and S. F. Hviid\inst{\ref{inst22}}
\and W.--H. Ip\inst{\ref{inst23}}
\and L. Jorda\inst{\ref{inst6}}
\and J. Knollenberg\inst{\ref{inst22}}
\and G. Kovacs\inst{\ref{inst3}} 
\and J.--R. Kramm\inst{\ref{inst3}} 
\and E. K\"uhrt\inst{\ref{inst22}}
\and M. K\"uppers\inst{\ref{inst24}}
\and L.M. Lara\inst{\ref{inst1}} 
\and M. Lazzarin\inst{\ref{inst4}} 
\and J. J. L\'opez--Moreno\inst{\ref{inst1}}
\and F. Marzari\inst{\ref{inst4}} 
\and S. Mottola\inst{\ref{inst22}}
\and G. Naletto\inst{\ref{inst5},\ref{inst17},\ref{inst25}}
\and N. Oklay\inst{\ref{inst3}} 
\and M. Pajola\inst{\ref{inst5}}
\and N. Thomas\inst{\ref{inst20}} 
\and J.B. Vincent\inst{\ref{inst3}}
\and V. Della Corte\inst{\ref{inst55}}
\and A. Fitzsimmons\inst{\ref{inst52}}
\and S. Faggi\inst{\ref{inst53}}
\and E. Jehin\inst{\ref{inst54}}
\and C. Opitom\inst{\ref{inst54}}
\and G.--P. Tozzi\inst{\ref{inst53}}
}
   \institute{
Instituto de Astrof\'\i sica de Andaluc\'\i a, CSIC,
     Glorieta de la Astronom\'\i a s/n, 18008 Granada, Spain\\
              \email{fernando@iaa.es}\label{inst1}
         \and
Planetary and Space Sciences, Department of Physical
Sciences, The Open University, Walton Hall, Milton Keynes,
Buckinghamshire MK7 6AA, UK\label{inst50}
\and
European Southern Observatory, Karl-Schwarschild-Strasse 2, D-85748
Garching bei Mu\"{u}nchen, Germany\label{inst51}
\and Max-Planck Institut f\"{u}r Sonnensystemforschung, Justus-von-Liebig-Weg, 3 37077 G\"{o}ttingen, 
Germany \label{inst3}
\and Department of Physics and Astronomy G. Galilei, University of Padova,
Vic. Osservatorio 3,  
35122 Padova, Italy \label{inst4}
\and Center of Studies and Activities for Space (CISAS) “G. Colombo”, University of Padova, 
Via Venezia 15, 35131 Padova, Italy \label{inst5}
\and Aix Marseille Universit\'e, CNRS, LAM (Laboratoire dAstro-physique de Marseille) 
UMR 7326, 13388, Marseille, France \label{inst6}
\and Centro de Astrobiologia (INTA-CSIC), European Space Agency (ESA), European Space Astronomy Centre 
(ESAC), P.O. Box 78, E-28691 Villanueva de la Ca\~nada, Madrid, Spain \label{inst7} 
\and International Space Science Institute, Hallerstrasse 6, 3012 Bern, Switzerland \label{inst8} 
\and Research and Scientific Support Department, European Space Agency, 2201 Noordwijk, 
The Netherlands  \label{inst9} 
\and PAS Space Research Center, Bartycka 18A, 00716 Warszawa, Poland \label{inst10} 
\and Institute for Geophysics and Extraterrestrial Physics, TU Braunschweig, 38106 Braunschweig, 
Germany \label{inst11}
\and Department for Astronomy, University of Maryland, College Park, MD 20742-2421, USA \label{inst12} 
\and LESIA, Observatoire de Paris, CNRS, UPMC Univ Paris 06, Univ. Paris-Diderot, 
5 Place J. Janssen,  92195 Meudon Pricipal Cedex, France \label{inst13} 
\and LATMOS, CNRS/UVSQ/IPSL, 11 Boulevard dAlembert, 78280 Guyancourt, France  \label{inst14} 
\and INAF Osservatorio Astronomico di Padova, vic. dell’Osservatorio 5, 35122 Padova, Italy  \label{inst15} 
\and Department of Physics and Astronomy, Uppsala University, Box 516, 75120 Uppsala, Sweden \label{inst16} 
\and CNR-IFN UOS Padova LUXOR, via Trasea 7, 35131 Padova, Italy \label{inst17}
\and Department of Industrial Engineering, University of Padova, Via Venezia 1, 35131 
Padova, Italy \label{inst18}
\and University of Trento, via Sommarive 9, 38123 Trento, Italy \label{inst19}
\and Physikalisches Institut, Sidlerstrasse 5, University of Bern, CH-3012 Bern, Switzerland  \label{inst20} 
\and INAF -- Osservatorio Astronomico di Trieste, via Tiepolo 11, 34143 Trieste, Italy \label{inst21}
\and Deutsches Zentrum f\"{u}r Luft- und Raumfahrt (DLR), Institut
f\"{u}r 
Planetenforschung, 
Rutherfordstrasse 2, 12489 Berlin, Germany \label{inst22} 
\and Institute for Space Science, National Central University, 32054 Chung-Li, Taiwan \label{inst23} 
\and ESA/ESAC, PO Box 78, 28691 Villanueva de la Ca\~nada, Spain \label{inst24} 
\and Department of Information Engineering, University of Padova, via Gradenigo 6/B, 35131 Padova, 
Italy \label{inst25}
\and Instituto Nacional de T\'ecnica Aeroespacial, 28850 Torrej\'on de Ardoz, Madrid, Spain \label{inst26}
\and Astrophysics Research Centre, School of Physics and Astronomy,
Queen's University 
Belfast, Belfast BT7 1NN, UK \label{inst52}
\and  INAF, Osservatorio Astrofisico di Arcetri, Largo E. Fermi 5,
I-50 125 Firenze, Italy\label{inst53} 
\and  Institut d'Astrophysique et de G\'{e}ophysique, Universit\'{e}
de Li\`{e}ge, 
Sart-Tilman, B-4000, Li\`{e}ge, Belgium
\label{inst54} 
\and  Institute for Space Astrophysics and Planetology (IAPS),
National Institute for AstroPhysics (INAF), Via Fosso del Cavaliere
100, 00133 Roma
\label{inst55} 
}
\date{Received \today; accepted  XX}

 
  \abstract
   {The ESA Rosetta spacecraft, currently orbiting around comet 67P,
     has already provided in situ measurements of the dust grain properties
     from several instruments, particularly OSIRIS and GIADA. We propose
      adding value to those measurements by combining them with
     ground-based observations of the dust tail to monitor the
     overall, time-dependent dust-production rate and size distribution.}
   {To constrain the dust grain properties, 
we take Rosetta OSIRIS and GIADA results into account, and combine
     OSIRIS data during the approach phase (from late April to early
     June 2014) with a 
     large data set of ground-based images that were acquired with the ESO 
     Very Large Telescope (VLT) from February to November 2014.}
   {A Monte Carlo dust tail code, which has already been used to characterise the dust
     environments of several comets and active asteroids, has been
     applied to retrieve the dust parameters. Key 
     properties of the grains (density, velocity, and size
     distribution) were obtained from Rosetta observations: these
     parameters were 
     used as input of the code to considerably reduce 
      the number of free parameters. In this way, 
    the overall dust mass-loss rate and its dependence on the 
     heliocentric distance could be obtained accurately.}
   {The dust parameters derived from the inner coma measurements by
     OSIRIS and GIADA and from
     distant imaging using VLT data are consistent, except for the power
     index of the size-distribution function, which is 
      $\alpha$=--3, instead of $\alpha$=--2, for grains
     smaller than 1 mm. This is possibly linked to the presence of
     fluffy aggregates in the coma. The onset of  cometary activity
     occurs at approximately 4.3 au, with a dust 
      production rate of 0.5 kg/s, 
     increasing up to 15 kg/s at 2.9 au. This implies a dust-to-gas mass
     ratio
     varying between 3.8 and 6.5 for the best-fit model when combined with 
     water-production rates from the MIRO experiment.}
   {}


\titlerunning{The dust environment of 67P}

   \maketitle
%

\section{Introduction}

   \begin{figure*}
   \centering
   \includegraphics[angle=-90,width=18cm]{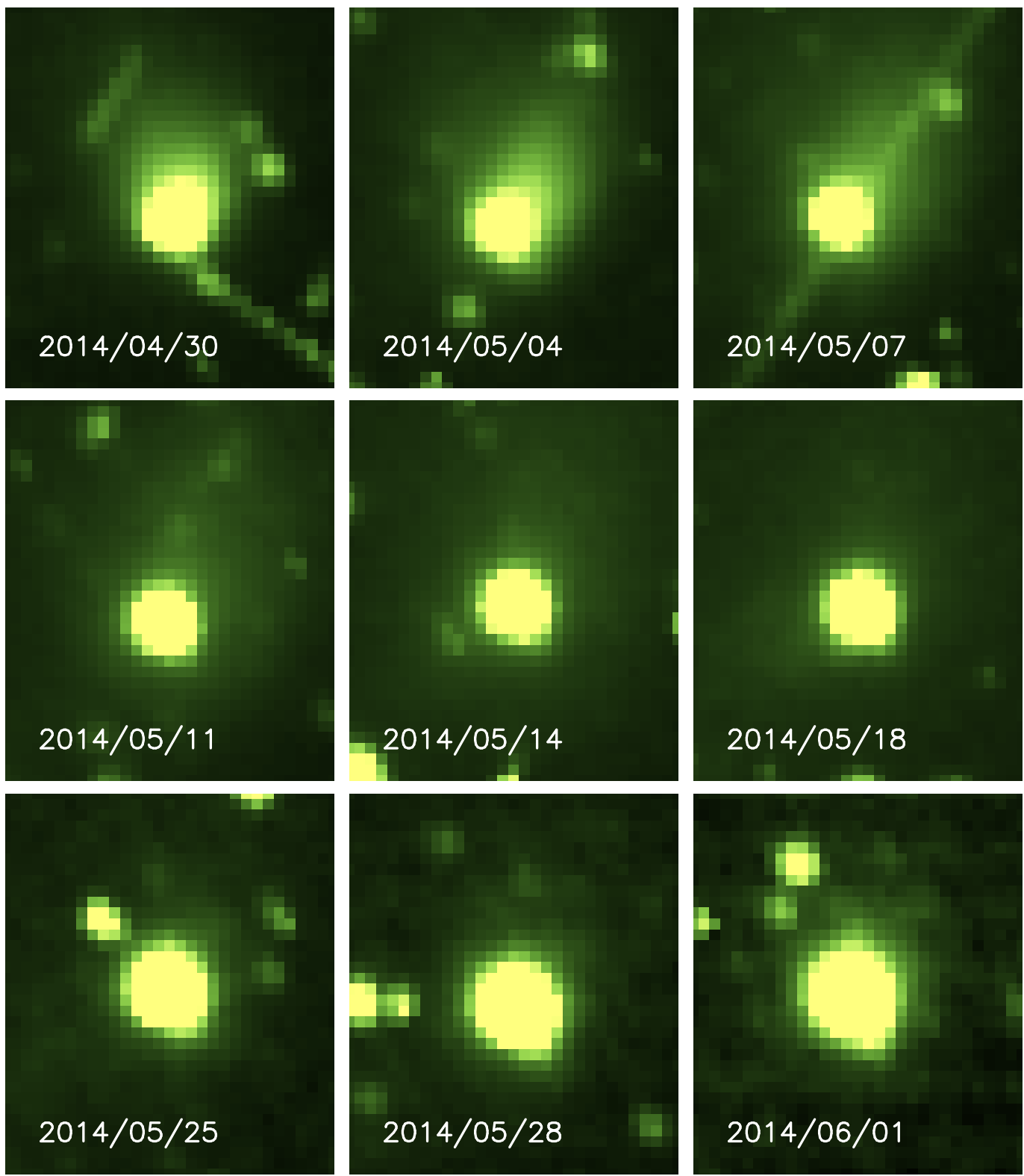}
   \caption{OSIRIS NAC images used in this work. All the images
     are oriented North up and East to the left. The lower labels in
     each image indicates the observation date. All images are
     30$\times$35 pixels. Their spatial dimensions at the comet 
     can be calculated
     using the spatial resolution indicated in the last column of
     Table~\ref{table1}. The straight features near the diagonals 
     in the uppermost first and third panels from the left are artifacts.}
              \label{FIG01}
    \end{figure*}

   Comets are among the least processed objects in the solar system, which means that   their study provides insights into their origin and evolution. The Grain
   Impact and Dust Accumulator (GIADA) \citep{Colangeli09,
     DellaCorte14} 
and the Optical, Spectroscopic,
   and Infrared Remote Imaging System (OSIRIS) \citep{Keller07}
   on board the ESA Rosetta spacecraft have been in operation in the
   vicinity of comet 67P/Churyumov-Gerasimenko (hereafter 67P) since March
   2014. The OSIRIS camera started to 
   provide images of the comet since March 23, 2014 \citep{Sierks15,Tubiana15a},
   while the first measurements carried out by GIADA
   were performed on July 18, 2014, when the comet was at heliocentric
   distances of 4.3 au and 3.7 au, respectively. 

Combining these OSIRIS and GIADA early 
   single grain measurements,  \citet{Rotundi15} derived the dust
   environment in the 3.7-3.4 au range, which  agrees with
   model predictions by \citet{Fulle10} at 3.2 au. Thus, the particle
   differential 
   size distribution was found to be adequately 
described by a differential power law of index  $\alpha$=--2, except
for the largest 
particle size bins ($r \gtrsim$1 mm) for which an index
$\alpha$=--4 is needed to satisfy the 67P trail data, which are mostly 
sensitive to those large particles, and which were acquired
in previous revolutions \citep{Agarwal10}. A close value of
$\alpha\sim$--3.7 for large particles 
has been found by \cite{Soja15} from analysis of 67P trail {\it
  Spitzer} data. This also agrees with the
distribution of blocks that are larger than a few centimetres in size 
on the surface of smooth terrains found by \cite{Mottola15} 
($\alpha$=--3.8$\pm$0.2) from the Rosetta Lander Imaging System
(ROLIS) on board Philae. This is also in line with the steep
distribution of large particles found by \cite{Kelley13a} in comet 
103P/Hartley 2 coma ($\alpha$=--4.7).  The maximum grain 
size ejected that was reported by 
\cite{Rotundi15} was 2 cm, and a dust loss rate
of 7$\pm$1 kg s$^{-1}$ was determined, giving a dust-to-gas mass ratio
(d/g) of 4$\pm$2 when combined with water production-rate data from
the Microwave Instrument for the Rosetta Orbiter (MIRO) 
\citep{Gulkis15}.  Assuming spherical particles, a grain density of 
1900$\pm$1100 kg m$^{-3}$ was derived.  Interestingly, the velocity
of outflowing grains did not show any link with particle
size. Hydrodynamic models of the inner coma \citep[e.g.][]{Crifo} 
would be needed to explain such behavior.  

Recently, a detailed study of GIADA data in the 3.4-2.3 AU heliocentric
distance range has been performed by \citet{DellaCorte15}. In this study, 
two populations of grains, both compact and fluffy, are described 
in relation to the geographical location of emission. Using the same
GIADA data, the  
contribution of the fluffy grain component to the brightness
of the coma has been found as being less than
15\% by \citet{Fulle15a}.  These fluffy grains have also been collected with the 
Cometary Secondary Ion Mass Analyser (COSIMA) on board Rosetta, 
which shows no trace of ice in their composition 
beyond 3 au \citep{Schulz15}.    

In contrast with the random distribution of velocities versus
grain size found by \citet{Rotundi15}, the relation 
of emission velocities and particle mass ($m$) found by \citet{DellaCorte15}
was quite steep, albeit
with a 50\% uncertainty, given 
by a power law $v \propto m^{\gamma_m}$ with $\gamma_m$=--0.32$\pm$0.18, which
translates to $v  \propto r^{\gamma_r}$ as 
$\gamma_r$=--0.96$\pm$0.54 ($r$ is particle radius). This is 
still compatible with 
the widely used $v  \propto r^{-0.5}$ dependence that is based on
simple hydrodynamical considerations \citep{Whipple50}. Photometric
measurements of individual grains  
at 2.25 AU and 2.0 AU from OSIRIS images suggest a $v
\propto r^{-0.5}$ dependence \citep{Fulle15inprep}. The different
dependencies of particle velocity from the particle radius determined by
\cite{Rotundi15} and \cite{DellaCorte15} might be
related to the various nucleus heliocentric distances and/or
different spacecraft-nucleus distances involved. Only
detailed hydrodynamic calculations in the inner coma, including the
detailed nucleus shape, will allow a proper interpretation of those
data. Consequently, the model runs with the different particle-velocity 
distributions since inputs
are needed to determine which distribution is the most 
compatible with our observations. 

   \begin{figure}
   \centering
   \includegraphics[angle=-90,width=\hsize]{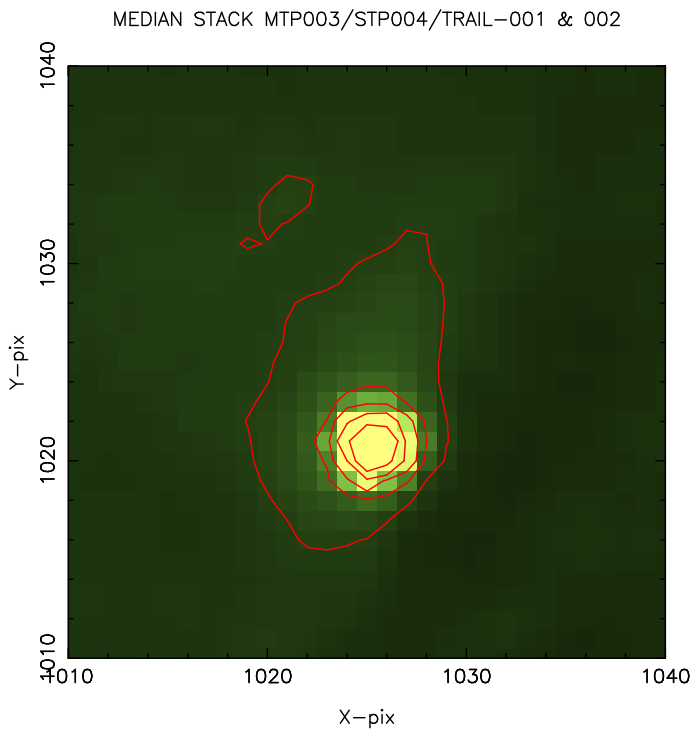}
   \caption{High signal-to-noise 
    WAC image obtained as a median stack of 117 frames,     obtained on May 17 and 18 2014. The dimensions of the image are
     3690$\times$3690 km at the nucleus distance. The innermost isophote level
     corresponds to 2.24$\times$10$^{-8}$ W m$^{-2}$ sr$^{-1}$ 
     nm$^{-1}$, decreasing outwards in factors of 2. North is up,
     East to the left.}
              \label{FIG02}
    \end{figure}

In this paper, we combine OSIRIS images acquired between 
April and July 2014 with ground-based images taken
with the FOcal Reducer and low dispersion Spectrograph 2 (FORS2) at 
the ESO VLT between February and November 2014, when the comet was
observable from the southern hemisphere. From late
November 2014 to late May 2015, the comet was behind the Sun for
ground observers. The comet 67P could be observed again from late May 2015 from the
southern hemisphere and from July 2015 from the northern
hemisphere. Ground-based observations, combined  
with the simultaneous OSIRIS data  described above, provide a 
unique opportunity to determine the dust properties and their time
evolution because they are most sensitive to the large-scale tail
structure that cannot be mapped from the spacecraft. Processes
such as significant volatile loss or grain fragmentation that take place a
few days after ejection cannot be monitored by OSIRIS.

   \begin{figure*}
   \centering
   \includegraphics[angle=-90,width=18cm]{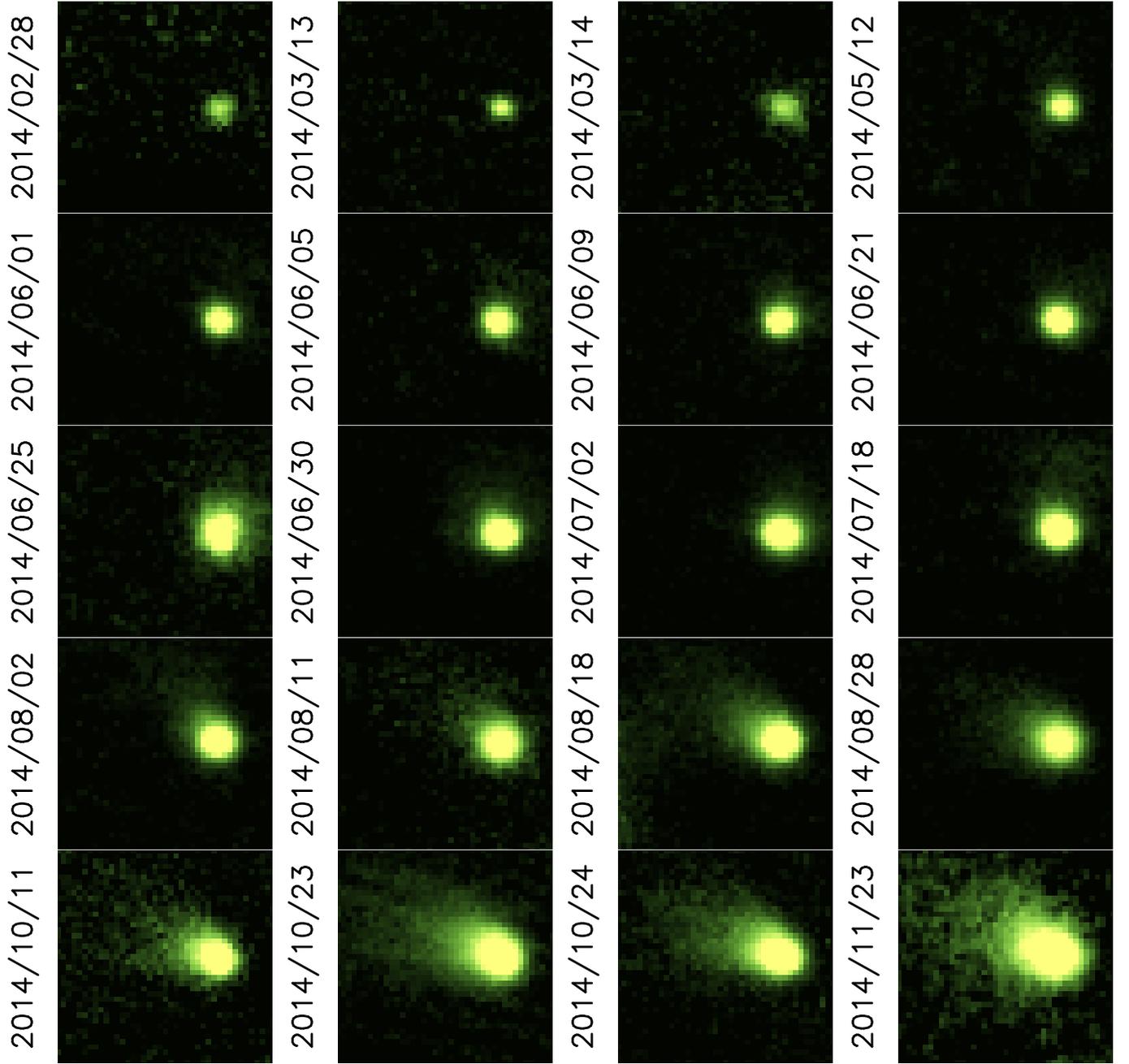}
   \caption{Subset of 20 out of the 52 VLT/FORS2 images taken between February
   and November 2014, with the R-SPECIAL filter. North is up, East
   to the left in all images. The date of observation is indicated in each
   panel. All the images have 40$\times$40 pixels in size, which can be
   converted to physical size at the comet using the spatial
   resolution values of last column of Table A.1 (see
   Appendix A).}
    \label{FIG03}
    \end{figure*}


\section{The Observations}

We used two data sets: 
\begin{itemize}
\item OSIRIS Narrow Angle Camera (NAC) and
Wide Angle Camera (WAC) images
obtained from late April to 
early July 2014. 
\item VLT/FORS2 images obtained from mid-February 
to late November 2014. 
\end{itemize}

A description of the instrumentation and the data reduction follows.

The OSIRIS instrument \citep{Keller07} comprises two cameras, the
NAC, with a field of view (FOV) of 2.20$^\circ\times$2.22$^\circ$
and an angular resolution of 1.87$\times$10$^{-5}$ rad px$^{-1}$,
and the WAC, with a FOV of 11.35$^\circ\times$12.11$^\circ$
with an angular resolution of 1.01$\times$10$^{-4}$ rad px$^{-1}$.  
We used NAC images taken with the Orange filter, which has a
central wavelength of 649.2 nm, and a bandwidth of 84.5 nm. The WAC
images used were taken with the Red filter, which has a central
wavelength of 629.8 nm and a bandwidth of 158.6 nm. In both cases, we
used Level 3 images, which were processed
following the standard OSIRIS calibration pipeline, including bias
subtraction, flat-fielding, distortion correction, and radiometric 
calibration \citep{Tubiana15b}. The log of the nine NAC observations
is given in  
Table~\ref{table1}, and Fig.~\ref{FIG01} displays the NAC
images. The WAC observations consisted of a sequence of 117 images
acquired 
between 2014-05-17T07:59:16 and 2014-05-18T06:52:05 (UT), and were aimed at
the detection of a trail feature. For the dust-tail modeling, we combined
all  117 frames into a single median stack, resulting in a high
signal-to-noise image (see Fig~\ref{FIG02}). The spatial
resolution of this image is 123 km px$^{-1}$.

\begin{table*}
\caption{Observational circumstances for the OSIRIS NAC observations}
\label{table1}      
\centering                          
\begin{tabular}{cccccr}        
\hline\hline                 
Date  & r  & S/C-Comet   & Phase angle  & Position angle & Resolution \\  
(UT) & (AU) & distance (AU) & (deg) & (deg) &(km px$^{-1}$) \\  
\hline                        
 2014-04-30T06.25.55.567 &      4.111 &     0.01586 &   35.10 & 120.697&  44.410 \\
 2014-05-04T03.17.55.567 &      4.093 &     0.01412 &   35.28 & 120.444&  39.557 \\
 2014-05-07T03.17.56.528 &      4.078 &     0.01278 &   35.40 & 120.234&  35.796 \\
 2014-05-11T12.49.46.645 &      4.057 &     0.01086 &   35.53 & 119.902&  30.410 \\
 2014-05-14T12.37.30.576 &      4.042 &     0.00956 &   35.56 & 119.647&  26.760 \\
 2014-05-18T12.21.10.575 &      4.023 &     0.00782 &   35.50 & 119.239&  21.896 \\
 2014-05-25T11.51.14.558 &      3.988 &     0.00540 &   35.32 & 118.497&  15.131 \\
 2014-05-28T11.38.14.574 &      3.973 &     0.00460 &   35.26 & 118.202&  12.891 \\
 2014-06-01T11.20.54.559 &      3.953 &     0.00354 &   34.95 & 117.682&   9.907 \\

\hline                                  
\end{tabular}
\end{table*}

The VLT data consist of CCD images
acquired with the 
R-SPECIAL filter (spectral response close to that of Bessell R, with
central wavelength of 655 nm and bandwidth of 165 nm). FORS was used
in imaging mode with the standard resolution collimator, and the
detector was read binning 2$\times$2, which resulted in an 
image scale of 0.25$\arcsec$ px$^{-1}$. The
reduction of the images was performed using standard techniques,
encompassing bias and flat-fielding, and photometric calibration by
standard star field imaging. In addition to the standard techniques, 
the images were also processed with a
background subtraction algorithm to remove the crowded star fields
\citep{Bramich08}. For each night of observation, a median
stack of the available images was produced and the magnitude 
within an aperture of 10 000 km radius was measured. Full details on
the data set and the reduction procedure are given by \cite{Snodgrass15}.  

The log of the observations is given in Table A.1 (see
Appendix A), and a subset of the
VLT images is displayed in Fig.~\ref{FIG03}. The images were
converted from mag arcsec$^{-2}$ ($m$) to mean solar disk intensity
units ($i/i_0$), according to the relationship  

   \begin{equation}
      m=2.5\log \Omega  + m_\sun -2.5\log (i/i_0)
   ,\end{equation}

where $\Omega$ is the solid angle of the Sun at 1 au
(2.893$\times$10$^6$ arcsec$^2$), and $m_\sun$ is the magnitude of the Sun
in the Bessell R filter ($m_\sun$=--27.09).

We
emphasise the importance of the different viewing angles and spatial scales of the
OSIRIS and VLT image sets. While
for the OSIRIS images the resolution ranges from 44 to 10 km
px$^{-1}$, the VLT images provide a spatial resolution from nearly
900 km px$^{-1}$ to $\sim$490 km px$^{-1}$ at the closest geocentric
distance of 2.7 au. 

An important aspect of the observations is the nucleus brightness 
contribution to the images. Since the OSIRIS images and the early VLT
images were  
acquired at large heliocentric distances when the comet was displaying 
little activity, the nucleus signal to the image
brightness dominates over the coma brightness and 
this must be taken into account. The total observed brightness  
of the images can 
be expressed as a sum of the nucleus plus the coma brightness
contributions convolved with the corresponding point spread function
(PSF) as \citep[e.g.][]{Lamy06}

    \begin{equation}
     B(x,y)=[C(x,y)+N(x,y)]\otimes P(x,y)
   ,\end{equation}

where $C(x,y)$, $N(x,y)$, and $P(x,y)$ are the coma, nucleus, and
PSF as functions of the pixel coordinates of the image relative to the
nucleus position, identified by the brightest pixel in the images.

To calculate the nucleus brightness at each epoch,
we developed a photon ray-tracing algorithm from which we constructed
synthetic images of the nucleus as it would be seen either from the
spacecraft or 
from the Earth. We used an early shape model of the nucleus (SHAP4S)
\citep{Preusker15} with a rotational axis pointing to RA=69.54$^\circ$,
DEC=64.11$^\circ$ (J2000 coordinates),  a Lambertian surface
model with a certain albedo value  
(the same for all the red filters used), a linear phase  
coefficient of 0.047 mag deg$^{-1}$ \citep{Fornasier15}, and a
rotational period of 12.4043h, the appropriate value at the time of
the observation \citep{Mottola14}. The albedo value was
constrained by matching a synthetic lightcurve to the 
experimental lightcurve derived from the sequence of WAC  
images obtained on May 17 and 18 2014, as described. The resulting
albedo is 0.06, which is close to the peak value of
0.063 obtained by \cite{Fornasier15} from the derived gaussian distribution of
surface albedos.

\begin{figure}
   \centering
   \includegraphics[angle=-90,width=\hsize]{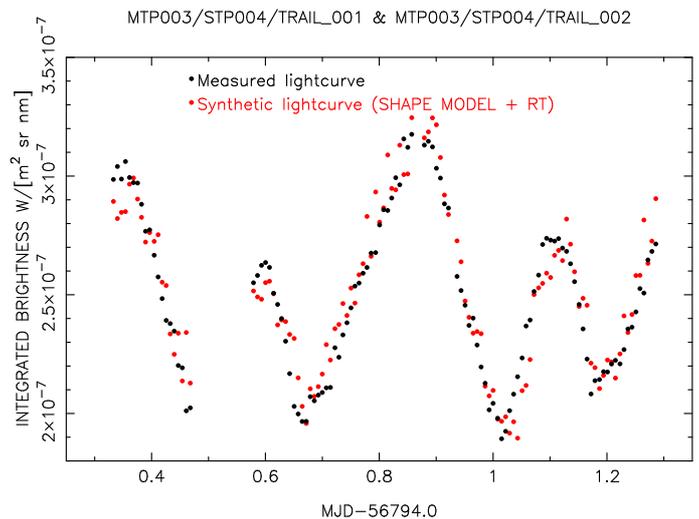}
   \caption{Lightcurve obtained from the analysis of 117 frames
     from WAC Red filter images obtained on 2014 May 17 and 18
     (black dots). The red dots correspond to a simulation of the
     lightcurve, taking in consideration  a nucleus shape model
     \citep{Preusker15} with a Lambertian model for the surface, which has an  
     albedo of 0.06, and a ray-tracing 
     technique to compute the total output flux at the S/C position.}
              \label{FIG04}
    \end{figure}

Although a noticeable coma is present in the images, 
its coma contribution to the
    integrated 
    nucleus plus coma system brightness at such heliocentric   
distances ($\sim$4 AU) is negligible, being estimated to be below the relative
photometric error of the measurements \citep{Mottola14}. This
contribution can be evaluated from the model results that are
described in the next sections. The integrated brightness for the WAC 
image of Fig.~\ref{FIG02} is 2.79$\times$10$^{-7}$ W m$^{-2}$
sr$^{-1}$ nm$^{-1}$. For our best-fit 
Model 1, which is characterised by a power-law particle-size distribution 
with power index of --3 and a random distribution of particle ejection 
velocities (see Sects. 3 and 4 for a detailed 
description of the model), we obtain 2.84$\times$10$^{-7}$ W m$^{-2}$
sr$^{-1}$ nm$^{-1}$ for the nucleus+coma system, which is in close agreement
with the measured brightness. From the same model, 
the integrated signal of the coma alone  is just
1.30$\times$10$^{-11}$ W m$^{-2}$ sr$^{-1}$ 
nm$^{-1}$, i.e. less than 0.005\% of the total signal. Thus, the coma
contribution can be safely ignored for the construction of the nucleus
lightcurve from the images at such heliocentric distances. To obtain the
integrated brightness at each observation date, an aperture
radius equal to the full width at half maximum (FWHM) of the field stars was 
used. Fig.~\ref{FIG04} shows a comparison of the measured lightcurve and the
one calculating the integrated brightness from the synthetic nucleus 
images, assuming a surface albedo of 0.06, as previously mentioned.
Along the lightcurve, the mean value of the difference between
measured and computed nucleus 
brightness is 1\%, while the maximum difference is 16\%. These
discrepancies are most likely related to the simplicity of the surface
model and to local variations of surface albedo \citep{Fornasier15}. 

The coma/tail brightness distribution is computed with the Monte Carlo
dust tail code. This is described in the next section.

\section{The model}

The Monte Carlo dust tail code has been described previously  
\cite[e.g.][]{Fulle10,Moreno12}. 
Briefly, this is a forward model that, given all the dust input
parameters, computes synthetic 
dust tail images for a given date by the trial-and-error procedure. To
this end, the code computes the  
trajectory of a large number of spherical particles that are ejected from a small-sized 
nucleus, assuming that the grains are affected by the solar
gravitation and the solar radiation pressure. The nucleus gravitational
field is neglected at  distances where the gas drag vanishes (at
approximately 20 nuclear radii, R$_N$). Under these assumptions, the
trajectory of the particles is Keplerian and the orbital 
elements are computed 
from the assumed terminal velocities (at 20R$_N$). The $1-\mu$ 
parameter \citep[e.g.][]{Fulle89}, which is the ratio of solar
radiation pressure force to solar gravity force, is 
given by $1-\mu =
C_{pr}Q_{pr}/(2\rho r)$. In this equation, $C_{pr}$ is given by

\begin{equation}
C_{pr} = \frac{3E_s}{8\pi cGM_{\sun} } = 1.19\times 10^{-3} \text{kg
 m$^{-2}$}
,\end{equation}

where $E_s$ is the mean total solar radiation, $c$ is the speed of
light, $G$ is the universal gravitational constant, and $M_\sun$ is
the solar mass \citep{Finson68}. $Q_{pr}$ is the radiation pressure
coefficient,  
and $\rho$ is the particle density. The radiation pressure coefficient for 
absorbing particles with radii $r \gtrsim$1 $\mu$m is $Q_{pr}\sim$1
\citep[e.g.][]{Moreno12}. The geometric albedo of the particles 
at red wavelengths is assumed to be $p_v$=0.065, a value close to to the one that was 
determined for the geometric albedo of the nucleus, and which ranged from
0.0589 to 0.072 at 535 and 700 nm, respectively, from disk-averaged photometry  
using OSIRIS NAC data \citep{Fornasier15}. For the dust particles, we assumed 
a phase angle coefficient of 0.03 mag 
deg$^{-1}$ in agreement with the value reported by \cite{Snodgrass13}
for comet 67P, and within the range estimated for other comets 
\citep[e.g.][]{MeechJewitt87}. In 
addition to the input model parameters just
described, we need to set the size-distribution function and the dust-loss rate, as a function of the heliocentric distance.  

\begin{figure}
   \centering
   \includegraphics[angle=-90,width=\hsize]{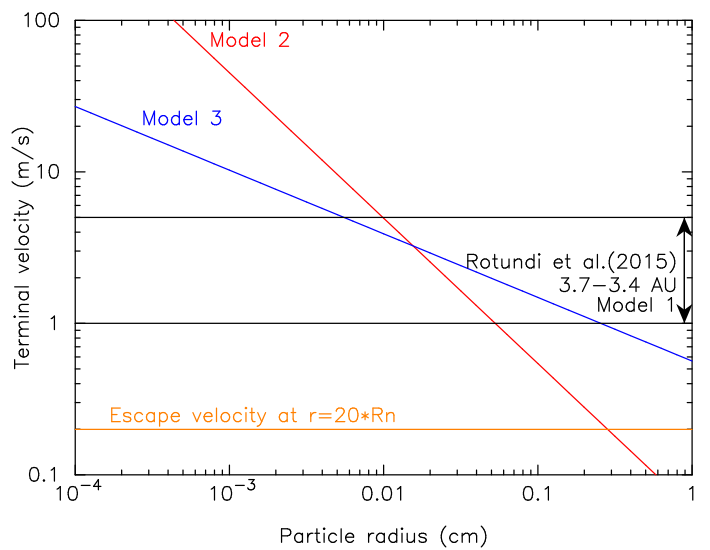}
   \caption{Size-dependent component of the terminal velocity of
     the dust grains assumed in the Monte Carlo model. Model 1
     corresponds to the random velocity distribution found by
     \cite{Rotundi15}, Model 2 corresponds to the steep, most probable
     value, 
     $v\propto r^{-0.96}$ found by \cite{DellaCorte15}, and Model
     3 to the less steep dependence given by $v\propto r^{-0.42}$. The escape
     velocity at 20R$_N$ is also indicated.}
              \label{FIG05}
    \end{figure}

\begin{figure}
   \centering
   \includegraphics[angle=-90,width=\hsize]{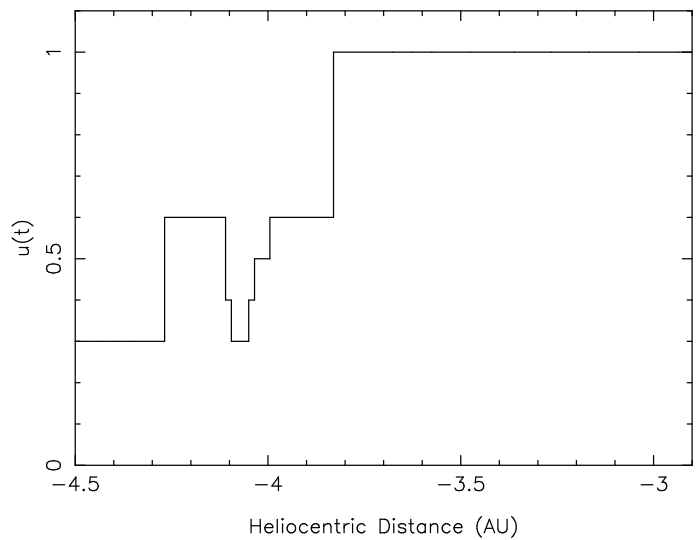}
   \caption{Assumed time-dependent component of the terminal velocity of
     the dust grains assumed in the Monte Carlo model. The bump
     between --4.25 to --4.15 AU is related to the 
     30 April 2014 outburst.}
              \label{FIG06}
    \end{figure}

In most previous applications of the code, 
the dust tail code was applied with almost no
previous knowledge of any of the dust properties. However, the 
observations of the inner coma dust grains by OSIRIS and GIADA
provide a unique characterisation of fundamental dust
parameters as described in Sect. 1. 
We used the dust physical parameters described by
\citet{Rotundi15} as inputs for the Monte Carlo code: the particle-size
distribution is governed by a power law with index --2 (except at
sizes larger than 1 mm, for which the index is --4), the 
largest particles ejected are 1 cm in radius, 
and the particle density is set to 2000 kg
m$^{-3}$. Taking into account the distributions of particle
velocities found by  \citet{Rotundi15} and \citet{DellaCorte15}, 
we have devised three different models with distinct grain-velocity dependencies, all of the form $v(\mu,t)=v_1(\mu)u(t)$,
$v_1(\mu)$ being a size-dependent velocity function, and $u(t)$
a dimensionless time-dependent parameter. These models range from a velocity that is independent of particle
size, as found by \citet{Rotundi15}, to a steep function of velocity on
grain size, as derived in \citet{DellaCorte15}. Specifically, in Model 1, the
terminal velocity of the grains is given by $v(\mu,t)=rnd[1,5] u(t)$, where
$rnd[1,5]$ is a random value in the 1 to
5 m s$^{-1}$ interval; in Model 2, the terminal velocity is given by
 $v(\mu,t)=1320(1-\mu)^{0.96} u(t)$ m s$^{-1}$; and in Model 3, we set  
$v(\mu,t)=45(1-\mu)^{0.42} u(t)$ m s$^{-1}$, in agreement with the
most probable value, and the lower limit in the power-law exponent of
the velocity versus particle size as determined by
\citet{DellaCorte15}, respectively. We note 
that these functions are, in all
cases, derived for a certain range of particle radii only, within the
range of instrumental sensitivity,  mostly in the
0.1 to 10 mm domain.  Fig.~\ref{FIG05} displays the just described
size-dependent term of the terminal velocities for each model 
and Fig.~\ref{FIG06} shows 
the time-dependent component $u(t)$.  

The function $u(t)$ reflects 
the overall time-dependence of the velocity and is constrained 
 as $u(t)=1$ in the heliocentric distance range 3.7-3.4
au from the results of \citet{Rotundi15} for Model 1. Out of this
heliocentric distance range, and for Models 2 and 3, a
free parameter  is adjusted by the trial and error
    procedure, taking into account that it should increase after the
    onset of activity \citep[e.g.][]{Fulle10}, and it is further
    enhanced whenever an outburst occurs, as has been observed, e.g. after
    the 2007 outburst of comet 17P/Holmes
    \citep{Montalto08,Moreno08}. The other  
free parameters that must be set are the dust-loss
rate as a function of the heliocentric distance, and the activation
time. The heliocentric dependence of the dust-mass loss rate is
    initially assumed as a function that varies as the inverse square of the
    heliocentric distance, and then several tens of test
    functions that modify the initial profile are  
    introduced until a good fit to all the  datasets of OSIRIS and VLT
    images is reached. Thus, the presence of outbursts, such as that of
    30 April 2014 \citep{Tubiana15a} imply a sudden increase of the
    dust-loss rate that clearly has to be  taken into account to fit
    the data. The goodness of the fit for each model is characterized
    by the quantity
\begin{equation}
\chi^2  =\left| \sum_{i,j} \frac{
  \left[\log(I_{obs}(i,j))-\log(I_{mod}(i,j))\right]^2 }
{  \log(I_{mod}(i,j))  } \right|,   
\end{equation}
where $I_{obs}$ and $I_{mod}$ are the observed and modeled brightness, and 
the summation is extended to the pixels $(i,j)$ along a scan 
in the anti-solar direction (position angle of the Sun-to-comet radius
vector), passing 
through the brightest pixel.  This $\chi^2$ parameter is 
calculated for each image, and then the total 
summation of the $\chi^2$ for all the OSIRIS plus VLT images is the
parameter used as the quality of the fit for each model. As
previously stated, the size distribution is initially set  to 
a power law of 
index --2 and the maximum particle radius to 1 cm, in accordance with
\citet{Rotundi15}. The minimum particle radius is set initially to 1
$\mu$m, but tests will also be made for smaller and larger minimum
size, as described below.

Another input parameter of the model is the emission
pattern. The OSIRIS images acquired in August 2014 
when Rosetta was at a distance from the comet, when the nucleus
shape could be clearly resolved  
\citep{Sierks15,Lara15}, revealed a number of conspicuous jets coming
from the neck region (Hapi) and pointing towards the direction of the
pole. In our model, we imposed a relatively higher number of
particles launched from latitudes higher than 60$^\circ$ than from
elsewhere. Specifically, we set 35\% of particles (in number) as being
emitted from that region and 65\% from latitudes south of
60$^\circ$. This corresponds to a flux of particles 
that are 7.5 times higher at those high latitudes than elsewhere. To
keep the number of free parameters to a minimum, we have not attempted
to modify the emission pattern along the orbital arc.  
This highly anisotropic
ejection pattern, with particle emission directed to the north, has
already  been  
inferred by \citet{Fulle10} in the GIADA dust environment model from
model fits to 67P data that was acquired during three revolutions around the
Sun prior to the Rosetta encounter.

\section{Results and discussion}

Once a set of model parameters has been defined, we run the Monte Carlo dust
tail code for all the OSIRIS and VLT images involved in the
analysis. However, for clarity, and to save space, we will only show
the results for subsets of them. Specifically, four dates are
selected for the OSIRIS images, and twenty for the VLT images. In
each case, the isophotes will be shown in the same brightness units of
W m$^{-2}$ sr$^{-1}$ nm$^{-1}$ (as for the Level 3 OSIRIS images), and
in units of the solar disk intensity for the VLT images. 

   \begin{figure}
   \centering
   \includegraphics[angle=-90,width=\hsize]{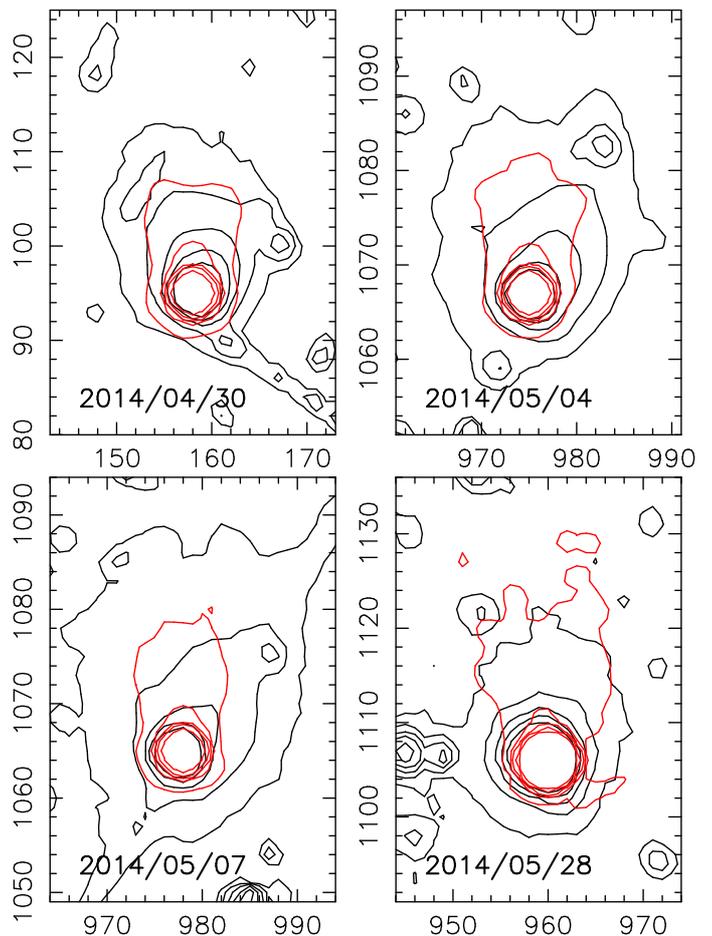}
   \caption{Monte Carlo dust tail code simulations (red contours)
     compared to the OSIRIS NAC observations (black contours) for the
   input parameters derived from \cite{Rotundi15} with a power-law
   exponent of the size distribution function of --2. The innermost
   isophotes have a value of 3.2$\times$10$^{-8}$  W m$^{-2}$
   sr$^{-1}$ nm$^{-1}$ 
   in all  four images and decrease in factors of 2 outwards.}
              \label{FIG07}
    \end{figure}

   \begin{figure}
   \centering
   \includegraphics[angle=-90,width=\hsize]{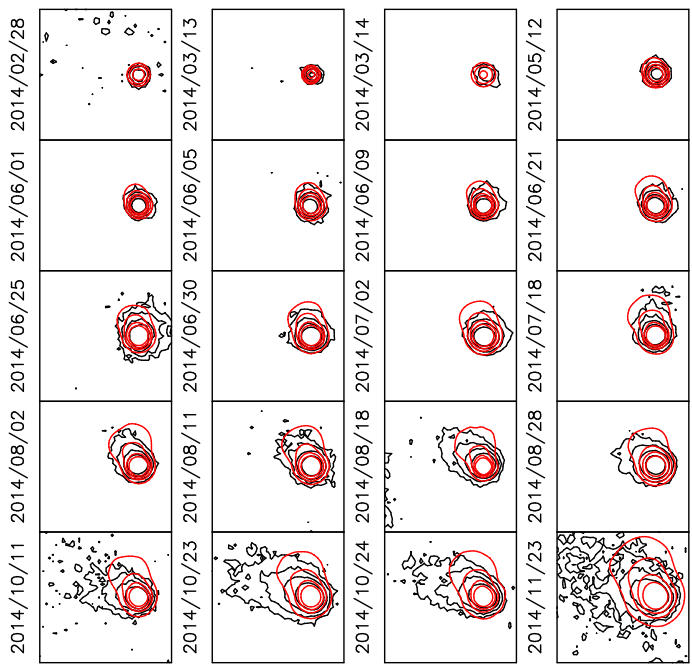}
   \caption{Monte Carlo dust tail code simulations (red contours)
     compared to the VLT observations (black contours) for the
   input parameters derived from \cite{Rotundi15} with a power-law
   exponent of the size distribution function of --2. The image dates
   are shown on the left side of each panel. Innermost isophotes are
   4$\times$10$^{-12}$ solar disk units, except for panels 2014/02/28 and
   2014/03/14, which have 2$\times$10$^{-12}$ solar disk units.}
              \label{FIG08}
    \end{figure}

\begin{figure}
   \centering
   \includegraphics[angle=-90,width=\hsize]{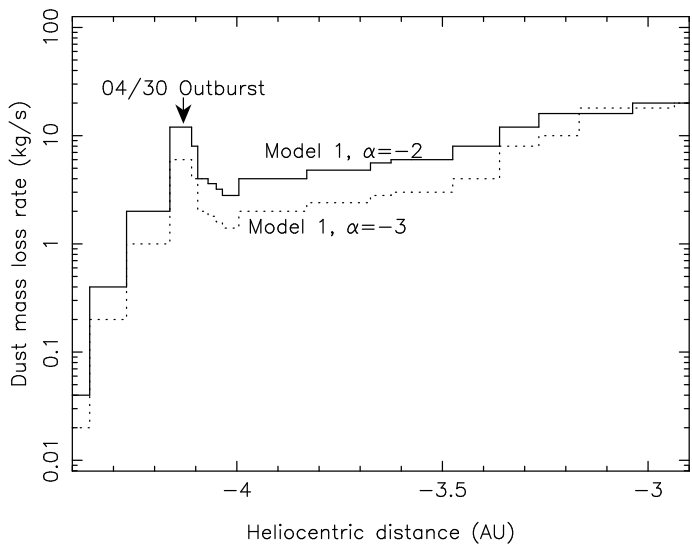}
   \caption{Best-fit dust-mass loss rate as a function of the
     heliocentric distance for power-law size distributions of --2
     (solid line) and --3 (dotted line). Note the increase in mass loss
     rate at the time of the 30 April 2014  outburst.}
              \label{FIG09}
    \end{figure}

We start by displaying the results of Model 1, the one which has a
random distribution of 
terminal particle velocities as described above. For this model, we
obtain the results shown in  Figs.~\ref{FIG07} and~\ref{FIG08} 
for the OSIRIS NAC
and VLT images, respectively, corresponding to the best-fit dust-mass
loss rate displayed in  Fig.~\ref{FIG09}. This graph shows a peak at
about 4.11 au pre-perihelion that  
corresponds to the outburst that occurred on 30 April 2014, in which 
a total dust-loss mass of $\sim$7$\times$10$^6$ kg was released. The
comet shows some activity, at the level of $\sim$0.5 kg
s$^{-1}$, already at 4.3 au pre-perihelion. This is in agreement with
the findings of \citet{Snodgrass13} for the previous orbit and
their predictions for the current orbit. In their trail model, 
\cite{Soja15} reported values of $\sim$10 
kg s$^{-1}$ at 3 au, which is in line with ours, but of $\sim$3.5 kg
s$^{-1}$ at 4.3 au, a value that is considerably larger than our estimate at that
heliocentric distance.

   \begin{figure}
   \centering
   \includegraphics[angle=-90,width=\hsize]{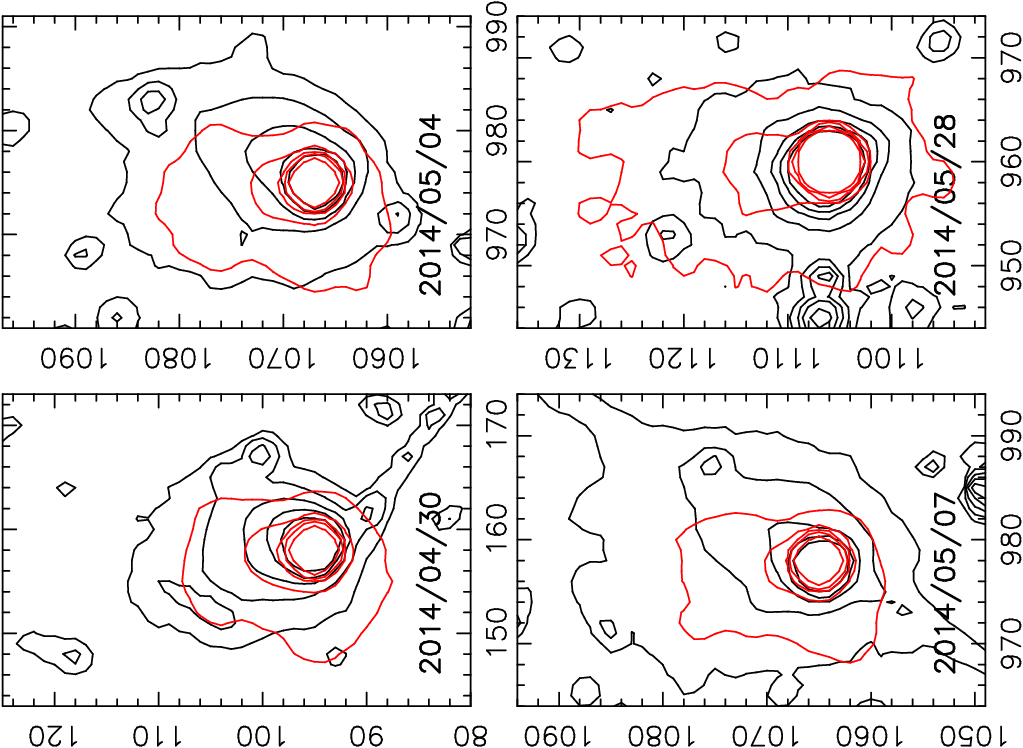}
   \caption{Monte Carlo dust tail code simulations (red contours)
     compared to the OSIRIS NAC observations (black contours) for the
   input parameters derived from \cite{Rotundi15} with a power-law
   exponent of the size distribution function of --3 and the random
   distribution of terminal velocities (Model 1).}
              \label{FIG10}
    \end{figure}

   \begin{figure}
   \centering
   \includegraphics[angle=-90,width=\hsize]{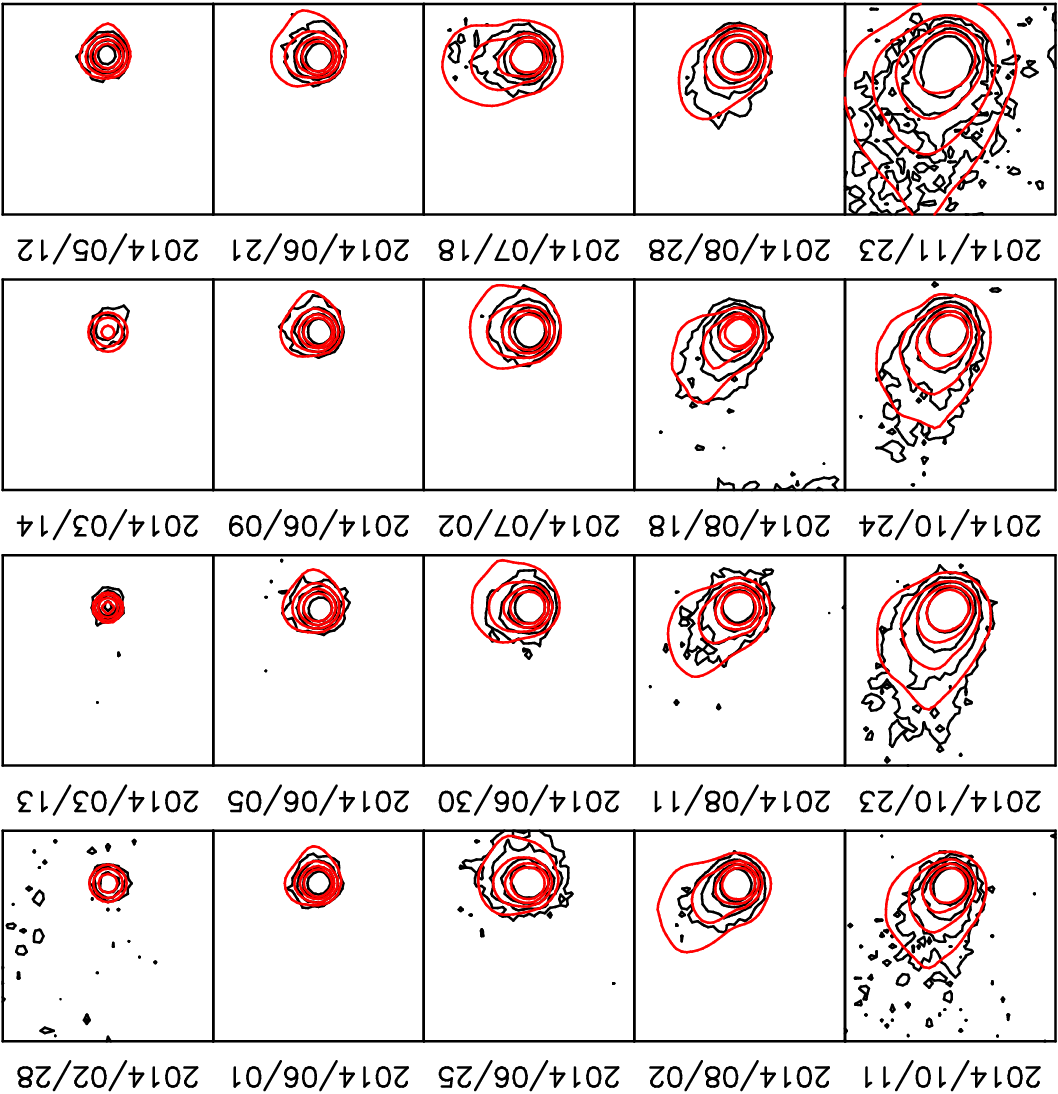}
   \caption{Monte Carlo dust tail code simulations (red contours)
     compared to the VLT observations (black contours), for the
   input parameters derived from \cite{Rotundi15}, with a power-law
   exponent of the size distribution function of --3, and the random
   distribution of terminal velocities (Model 1).}
              \label{FIG11}
    \end{figure}

While the modelled images fit the tail
shapes for OSIRIS and VLT images until early August 2014 reasonably well, it 
is clear that they 
deviate from the measured isophotes at later dates. The total $\chi^2$
  for this model is 40.1. No improvements are found when Models 
2 and 3 are run with different velocity distributions. We therefore tried to modify the
input parameters to obtain a better fit to the
data. We found that no improvements were
possible, except when the power-law index of the size distribution
is decreased, i.e. using a steeper size-distribution function. The best
results were found for a power index of --3, for which the
model results are presented in  Figs.~\ref{FIG10} and  \ref{FIG11} 
for the NAC
and the VLT images, respectively, for Model 1. The corresponding
best-fitted dust-loss rate is shown in  Fig.~\ref{FIG09}. For
    this model,  $\chi^2$ is 14.4, whcih is much smaller than the value of
    40.1 found for the model with power index of --2. 

The best-fit power index of --3 for the size-distribution function 
agrees with that previously derived by \citet{Fulle10}, 
which was obtained from dust tail modelling of observations that were acquired during several
previous orbits. As noted by \citet{Fulle15a}, the different power
indices of  --2 derived by 
\citet{Rotundi15} and of --3 derived from the dust tail models,
could be related to contribution of the aggregate particles. The presence of 
fluffy grains in the coma has been 
clearly shown by the GIADA instrument since mid-September 2014 
\citep{DellaCorte15,Fulle15a}. As \citet{Rotundi15} use 
earlier measurements,  
the power index of --2 only refers to the compact particle population. It is possible that those aggregates, which have a 
steeper size distribution function, are contributing to solve this
inconsistency \citep{Fulle15a}. However, proving it quantitatively is far from
easy. First, the computation of the $(1-\mu)$ parameter for
aggregates of the order of, or larger 
than, the wavelength of the incident light 
is prohibitive in terms of current computational resources. And
second, the 
non-radial pressure force component on the aggregates becomes
non-negligible \citep{Kimura02}, implying a non-Keplerian motion for
those particles, and the use of numerical integrators to determine
their orbits. In any case, it is interesting to note that
the later the observation date,  
the larger the discrepancy is between the models with size-distribution power indices of --2 and --3  (compare Fig.~\ref{FIG08} and
Fig.~\ref{FIG11}). 

\begin{figure}
   \centering
   \includegraphics[angle=-90,width=\hsize]{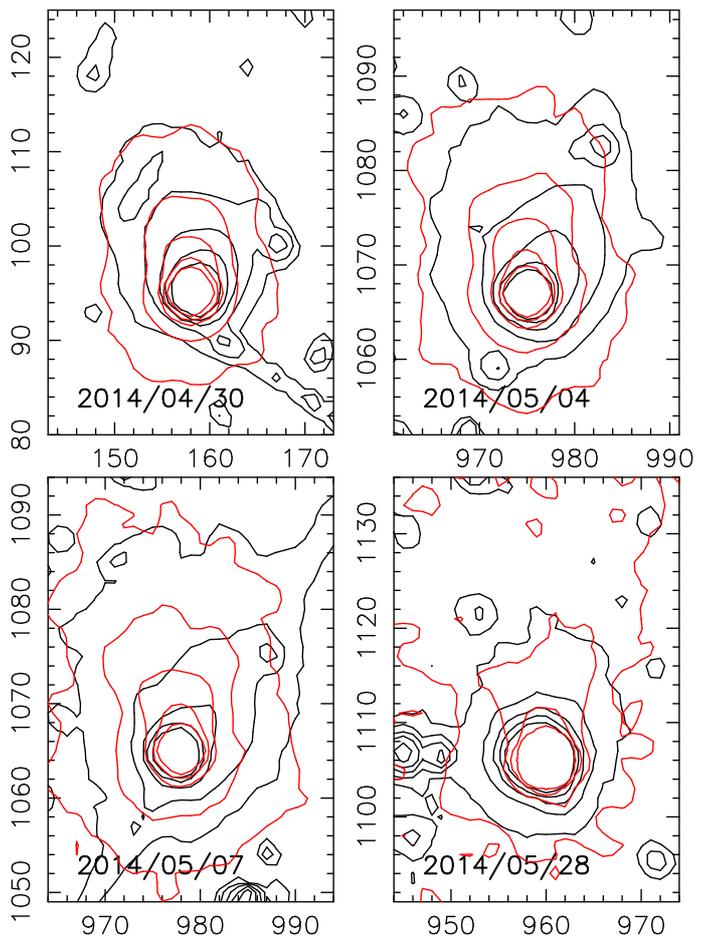}
   \caption{Monte Carlo dust tail code simulations (red contours)
     compared to the OSIRIS NAC observations (black contours) for the
   input parameters derived from \cite{Rotundi15} with a power-law
   exponent of the size distribution function of --3 and the terminal
   velocities being given by $v(\mu,t)=1320(1-\mu)^{0.96} u(t)$ m s$^{-1}$ 
   (Model 2).}
              \label{FIG12}
    \end{figure}

\begin{figure}
   \centering
   \includegraphics[angle=-90,width=\hsize]{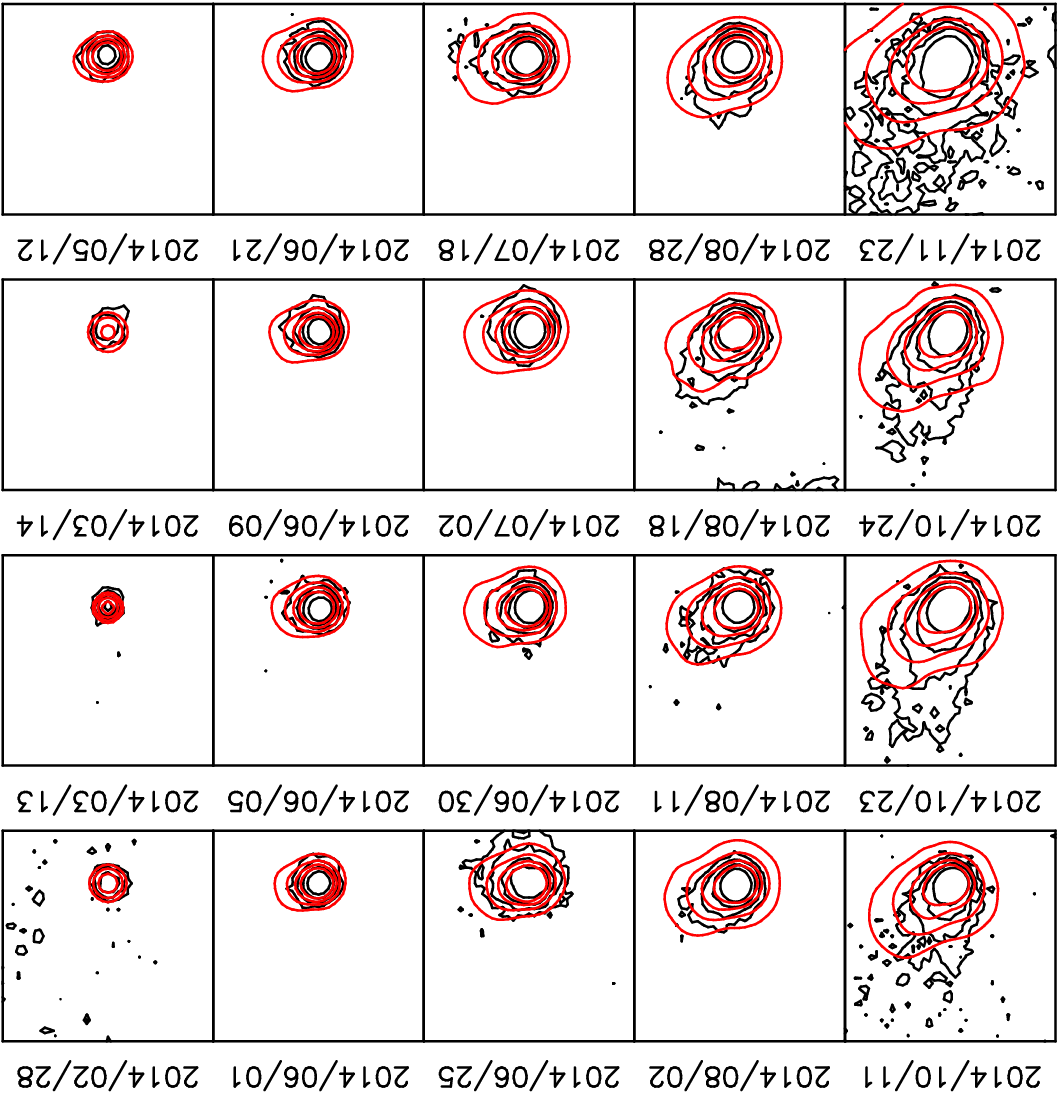}
   \caption{Monte Carlo dust tail code simulations (red contours)
     compared to the VLT observations (black contours) for the
   input parameters derived from \cite{Rotundi15} with a power-law
   exponent of the size distribution function of --3 and the terminal
   velocities being given by  $v(\mu,t)=1320(1-\mu)^{0.96} u(t)$ m s$^{-1}$ 
   (Model 2).}
              \label{FIG13}
    \end{figure}

\begin{figure}
   \centering
   \includegraphics[angle=-90,width=\hsize]{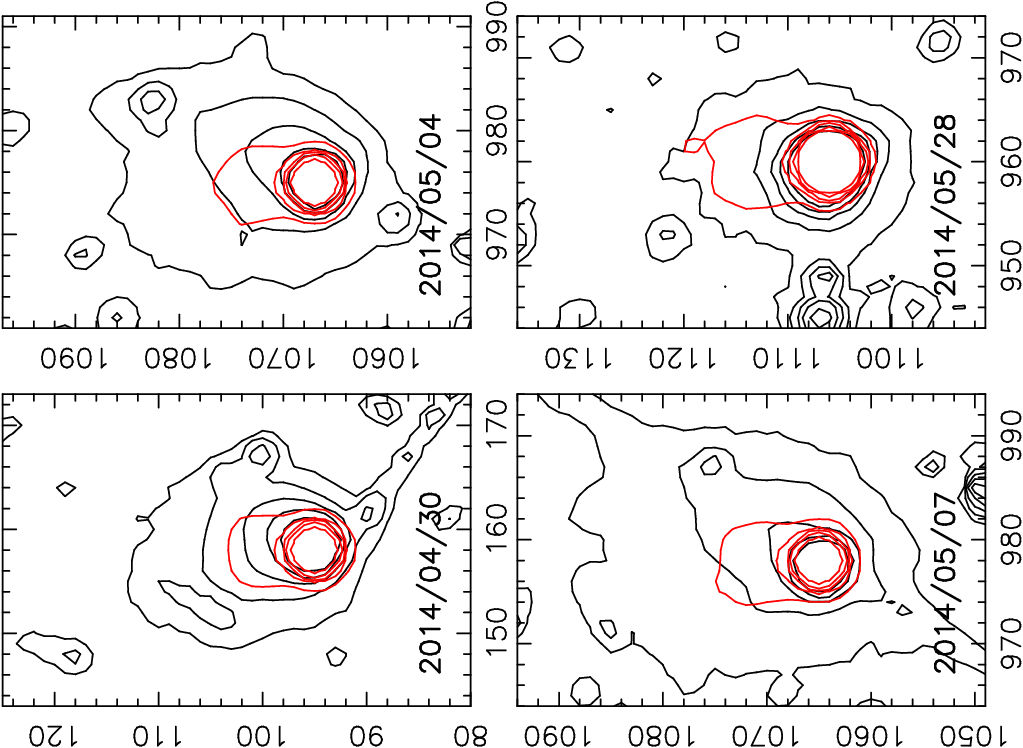}
   \caption{Monte Carlo dust tail code simulations (red contours)
     compared to the OSIRIS NAC observations (black contours) for the
   input parameters derived from \cite{Rotundi15} with a power-law
   exponent of the size distribution function of --3 and the terminal
   velocities being given by $v(\mu,t)=45(1-\mu)^{0.42} u(t)$ m s$^{-1}$ 
   (Model 3).}
              \label{FIG14}
    \end{figure}

\begin{figure}
   \centering
   \includegraphics[angle=-90,width=\hsize]{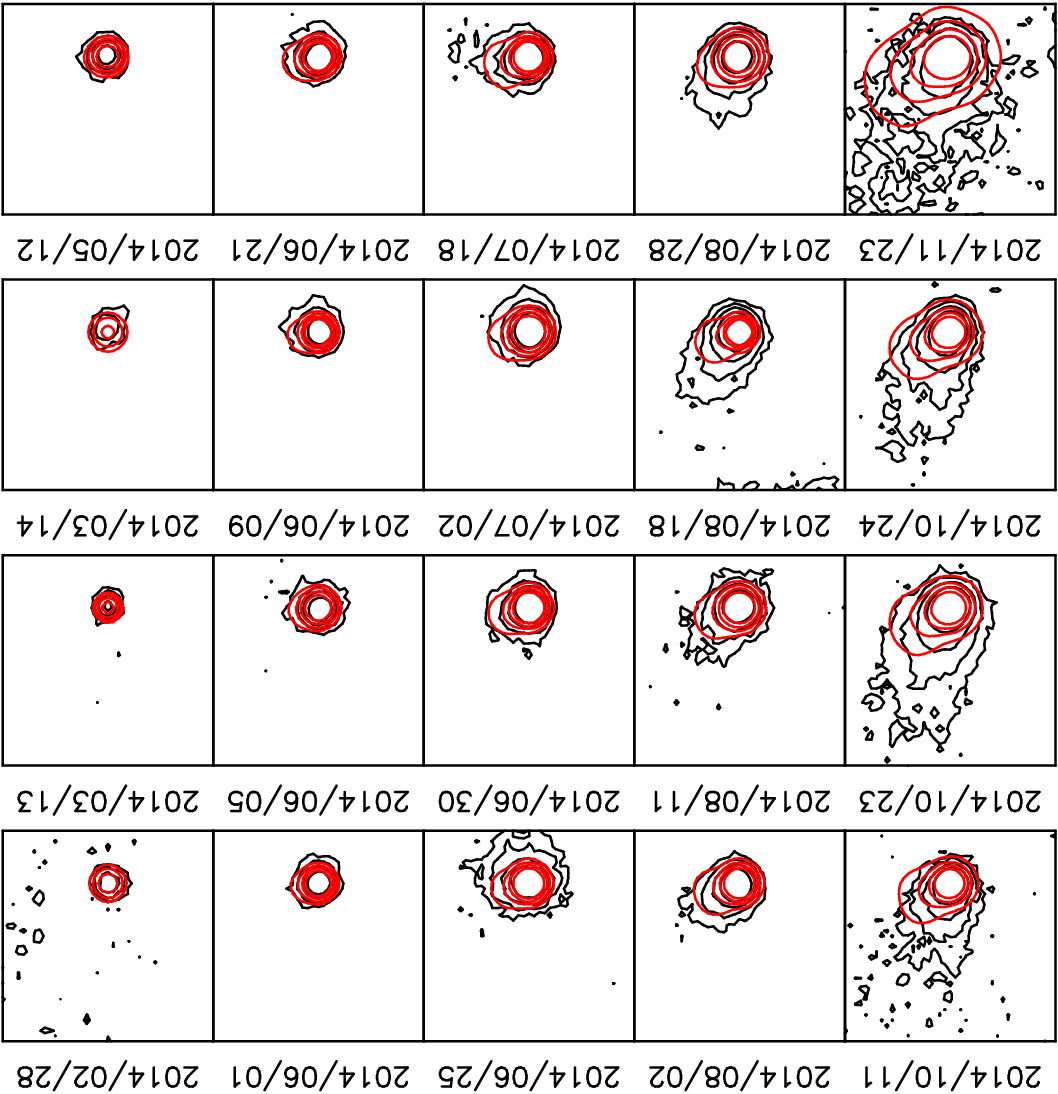}
   \caption{Monte Carlo dust tail code simulations (red contours)
     compared to the VLT observations (black contours) for the
   input parameters derived from \cite{Rotundi15}, with a power-law
   exponent of the size distribution function of --3, and the terminal
   velocities being given by $v(\mu,t)=45(1-\mu)^{0.42} u(t)$ m s$^{-1}$ 
   (Model 3).}
              \label{FIG15}
    \end{figure}

\begin{figure}
   \centering
   \includegraphics[angle=-90,width=\hsize]{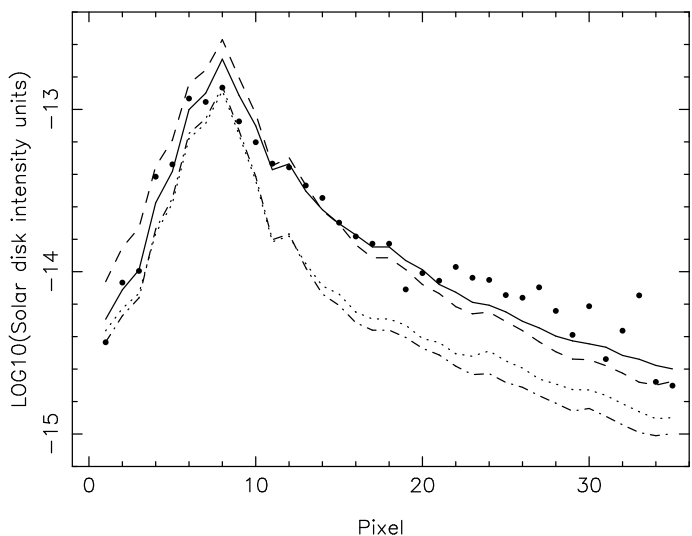}
   \caption{Scan along the anti-solar direction 
      of the VLT image of 2014/10/24 (solid circles) compared to
     the different ejection velocity models described in the text:      
     Model 1 (best-fit model) 
corresponds to the solid-line, Model 2 is represented by the dashed line, 
and Model 3 by the dotted line. These three models have  a power
index for the size distribution of --3. The dash-dotted line
corresponds to Model 1 but with a power index for the size 
distribution of --2.}
              \label{FIG16}
    \end{figure}

\begin{figure}
   \centering
   \includegraphics[angle=-90,width=\hsize]{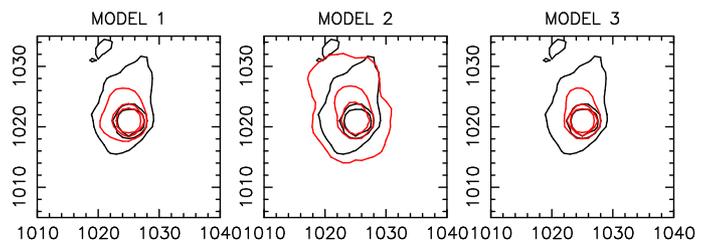}
   \caption{Monte Carlo dust tail code simulations (red contours)
     compared to the OSIRIS WAC observations (black contours), from
     left to right, for 
     Models 1, 2, and 3, with a power-law
   exponent of the size distribution function of --3.}
              \label{FIG17}
    \end{figure}

\begin{figure}
   \centering
   \includegraphics[angle=-90,width=\hsize]{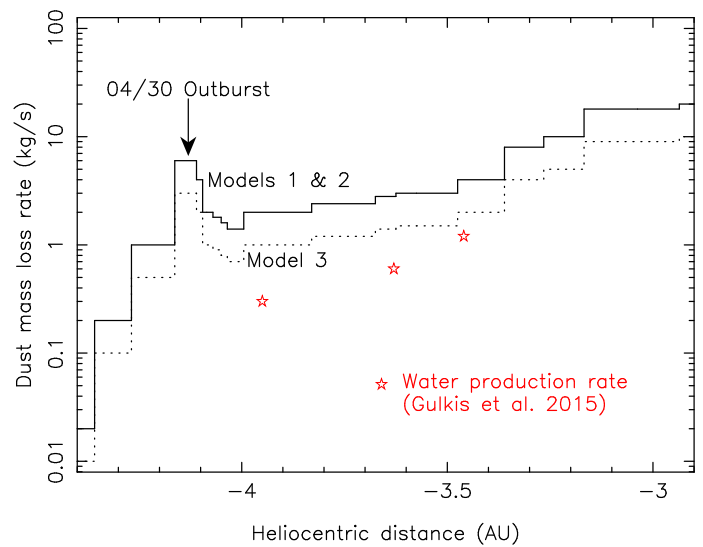}
   \caption{Best-fit dust-mass loss rate as a function of the
     heliocentric distance for Models 1 and 2 (solid line), and for
     Model 3 (dotted line), for 
     power-law size distributions of index --3. Also shown are the
     water production rates derived from MIRO instrument \citep{Gulkis15}.}
              \label{FIG18}
    \end{figure}

After finding the power index of --3 as the best fit for Model 1, we   
fit the data with Models 2 
and 3, assuming $\alpha$=--3 and leaving the dust-mass loss rate as
the only free 
parameter. The fits to the OSIRIS NAC and VLT data are found in
 Fig.~\ref{FIG12} and  Fig.~\ref{FIG13} for Model 2, and
 Fig.~\ref{FIG14} and  Fig.~\ref{FIG15} for Model 3. The 
models 1, 2, and 3 for the WAC data are shown in  Fig.~\ref{FIG17}, while
the corresponding best-fitted dust-loss rate functions are displayed
in  Fig.~\ref{FIG18}. Again,
a significant deviation of the model 
isophotes from the 
observed tail shapes is observed for the VLT data after early
August, and particularly for Model 3, 
which indicates that the particle terminal velocities are better
represented by a flat random distribution of velocities
rather than a power law of the size, as was found by
\citet{Rotundi15}. A proper
interpretation of the different particle velocities that were encountered by
\citet{Rotundi15} and \citet{DellaCorte15} should incorporate the
hydrodynamical processes in the inner coma, taking into account the
different ranges of nucleus to Sun and nucleus to spacecraft
distances involved. The computed values of $\chi^2$ are 16.1 and
    32.8 for models 2 and 3, respectively. As an example,
    Fig.~\ref{FIG16} depicts the resulting scans along the anti-solar 
    direction for one of the 
    latest VLT-observed and modeled images,
    where the deviations of the data for the different models can be seen.

 In Fig.~\ref{FIG18}, the water production rates at
different heliocentric distances 
that were obtained by \citet{Gulkis15} from MIRO are also plotted. From these
measurements, we obtain a d/g mass ratio varying between 6.5 at
3.95 au pre-perihelion and 3.8 at 3.5 au pre-perihelion, for Models 1 and 2, 
and between 3.3 at
3.95 au pre-perihelion and 1.6 at 3.5 au pre-perihelion 
for Model 3, which confirms the d/g=4$\pm$2 
that was estimated by \citet{Rotundi15}.

After obtaining this satisfactory fit for Model 1, we explored the sensitivity to
the input parameters of the model results. Changing the minimum 
particle radius from 1 $\mu$m to 10 $\mu$m does not significantly affect
 the $\chi^2$ of the fit,   
but if the minimum radius is set as high as 50 
$\mu$m, $\chi^2$ increases by up to 48.2. The
presence of dust particles smaller than 5 $\mu$m in size has been
demonstrated by \cite{DellaCorte15} from dust accumulation on two of the
five micro balances system (MBS) of GIADA since September, 2014.  On the
other hand, the input dust-mass loss rate can be varied up to the 
30\% level without altering the model results 
substantially. Thus, for Model 1, $\chi^2$ increases from 14.4
(best fit) to 16.2 by a variation of 30\% in the overall  
dust-mass loss rate. If the dust-loss rate profile is varied up to 50\%,  
then $\chi^2$ increases to 19.2, which is too large compared to its 
equivalent in the best fit
model. From this, we estimate a 50\% uncertainty  
as an upper limit to the obtained dust-loss rates.

Regarding the level of activity of the comet at the heliocentric
distance range explored, we compare the derived dust-loss
rates with other published estimates. In most cases, unfortunately,
only values of the quantity $Af\rho$ \citep{AHearn} have been
reported \citep[e.g.][and references therein]{Mazzotta09,Kelley13b}, 
which cannot be easily converted to dust-loss rates. \cite{Lamy09}
gave estimates of a few short-period comets at distances near 3 au,
namely 4P/Faye, 17P/Holmes, 44P/Reinmuth, and 71P/Clark. For those
comets, the dust-production rate near 3 au ranges from $\sim$1 to 5 kg
s$^{-1}$, which is less than that derived for 67P ($\sim$15 kg
s$^{-1}$).

\section{Conclusions}

A series of Rosetta OSIRIS NAC and WAC images, together with an extensive
data set of ground-based VLT images of comet 67P, have been analyzed
by our Monte Carlo dust tail model. At the time of the observations,
the comet moved from 4.4 au to 2.9 au, inbound. In this heliocentric distance
range, the main results can be summarised as follows:   

   \begin{enumerate}
      \item The comet was already active at 4.3 au,
        with a dust-mass loss rate of $\sim$0.5 kg s$^{-1}$, increasing up
        to $\sim$ 15 kg s$^{-1}$ at 2.9 au.  Based on the model
          results, 
these loss rates are affected by a maximum overall uncertainty of
$\sim$50\%.
      \item The dust size distribution is characterised by a power law
        with  a power index of --3 for particles smaller than 1 mm, and
        --4 for larger grains.
      \item The terminal velocity of the particles is best-fitted 
        by a random distribution in the 1 to 5 m s$^{-1}$ interval, 
        which is consistent with the in situ measurements of \cite{Rotundi15}.
      \item The minimum particle size is constrained at radius $r <$
        10 $\mu$m, while the maximum size is compatible with the
        innermost coma values derived by \cite{Rotundi15} ($r$=1 cm).
      \item The dust-to-gas mass ratio, based on the results of our
        best-fit model  and   
        combined with water production rates 
        that were inferred from the MIRO experiment, varies between 
        3.8 and 6.5, which is consistent with  
        the value of 4$\pm$2 derived by \cite{Rotundi15}.          
   \end{enumerate}
       
\begin{acknowledgements}

We are grateful to the anonymous reviewer for his/her 
comments and suggestions that helped to considerably improve  
the manuscript. \\

OSIRIS was built by a consortium led by the Max-Planck Institut f\"ur Sonnensystemforschung, G\"ottingen, Germany, in
collaboration 
with CISAS, University of Padova, Italy, the Laboratoire d’Astrophysique
de Marseille, France, the Instituto de Astrof\'\i sica de Andaluc\'\i
a, CSIC, Granada,
Spain, the Scientific Support Office of the European Space Agency, Noordwijk,
The Netherlands, the Instituto Nacional de T\'ecnica Aeroespacial,
Madrid, Spain, 
the Universidad Polit\'ecnica de Madrid, Spain, the Department of Physics and
Astronomy of Uppsala University, Sweden, and the Institut f\"ur Datentechnik und
Kommunikationsnetze der Technischen Universit\"{a}t Braunschweig, Germany.
The support of the national funding agencies of Germany (DLR), France
(CNES), Italy (ASI), Spain (MEC), Sweden (SNSB), and the ESA Technical
Directorate is gratefully acknowledged. We thank the Rosetta Science Ground
Segment at ESAC, the Rosetta Mission Operations Centre at ESOC and the
Rosetta Project at ESTEC for their outstanding work that enables the
scientific return
of the Rosetta Mission. \\

This paper is based on observations made with ESO telescopes at the La
Silla Paranal Observatory under programmes IDs 592.C-0924, 093.C-0593,
and 094.C-0054.

\end{acknowledgements}

\Online

\begin{appendix}

\section{Log of VLT/FORS2 observations}

\begin{table*}
\caption{Observational circumstances for the VLT/FORS2 observations}
\label{table2}      
\centering                          
\begin{tabular}{rccrrc}        
\hline\hline                 
Date  & r  & Earth-comet  & Phase angle  & Position angle & Resolution  \\  
dd.dd mm yyyy (UT) &  (au) & distance (au) & (deg) & (deg) &(km px$^{-1}$) \\  
\hline                        
  28.39  2 2014 &   4.386 &    4.909 &    10.40 &   259.27 &   890.1 \\
  13.37  3 2014 &   4.330 &    4.680 &    11.86 &   264.37 &   848.6 \\
  14.37  3 2014 &   4.326 &    4.661 &    11.96 &   259.27 &   845.2 \\  
  15.34  3 2014 &   4.321 &    4.643 &    12.06 &   259.27 &   841.9 \\  
  10.41  4 2014 &   4.204 &    4.133 &    13.77 &   259.65 &   749.5 \\
   4.37  5 2014 &   4.092 &    3.655 &    13.50 &   261.28 &   662.7 \\
   7.26  5 2014 &   4.078 &    3.599 &    13.32 &   261.61 &   652.7 \\
  12.40  5 2014 &   4.053 &    3.502 &    12.91 &   262.32 &   635.0 \\
   1.32  6 2014 &   3.954 &    3.159 &    10.22 &   266.94 &   572.8 \\
   5.12  6 2014 &   3.934 &    3.101 &     9.51 &   268.37 &   562.3 \\  
   6.20  6 2014 &   3.929 &    3.085 &     9.30 &   268.82 &   559.5 \\  
   9.32  6 2014 &   3.913 &    3.041 &     8.65 &   270.29 &   551.4 \\  
  10.22  6 2014 &   3.908 &    3.029 &     8.46 &   270.76 &   549.2 \\
  19.37  6 2014 &   3.861 &    2.915 &     6.34 &   277.39 &   528.5 \\  
  20.28  6 2014 &   3.856 &    2.904 &     6.11 &   278.33 &   526.6 \\  
  21.17  6 2014 &   3.851 &    2.895 &     5.89 &   279.30 &   524.9 \\  
  25.05  6 2014 &   3.830 &    2.855 &     4.92 &   284.60 &   517.6 \\  
  30.20  6 2014 &   3.804 &    2.808 &     3.64 &   296.05 &   509.2 \\  
   1.13  7 2014 &   3.799 &    2.801 &     3.43 &   298.98 &   507.9 \\
   2.09  7 2014 &   3.793 &    2.793 &     3.21 &   302.45 &   506.5 \\  
   7.05  7 2014 &   3.767 &    2.759 &     2.35 &   329.65 &   500.2 \\  
  15.07  7 2014 &   3.723 &    2.719 &     2.77 &    29.61 &   492.9 \\  
  16.08  7 2014 &   3.718 &    2.715 &     2.98 &    34.69 &   492.3 \\
  18.10  7 2014 &   3.707 &    2.708 &     3.44 &    42.98 &   491.1 \\  
  21.03  7 2014 &   3.690 &    2.701 &     4.20 &    51.65 &   489.8 \\  
  22.01  7 2014 &   3.685 &    2.699 &     4.47 &    53.92 &   489.4 \\  
  24.24  7 2014 &   3.673 &    2.696 &     5.09 &    58.24 &   488.9 \\  
  26.23  7 2014 &   3.662 &    2.694 &     5.66 &    61.35 &   488.6 \\
   2.21  8 2014 &   3.622 &    2.697 &     7.66 &    68.88 &   489.1 \\  
   4.16  8 2014 &   3.611 &    2.701 &     8.21 &    70.40 &   489.7 \\  
  11.99  8 2014 &   3.567 &    2.723 &    10.32 &    75.05 &   493.7 \\  
  15.99  8 2014 &   3.544 &    2.739 &    11.33 &    76.82 &   496.7 \\  
  16.98  8 2014 &   3.538 &    2.744 &    11.57 &    77.22 &   497.5 \\  
  18.00  8 2014 &   3.532 &    2.749 &    11.81 &    77.61 &   498.4 \\
  28.98  8 2014 &   3.467 &    2.814 &    14.16 &    80.99 &   510.2 \\
  23.16  9 2014 &   3.315 &    3.016 &    17.42 &    85.11 &   546.9 \\
  11.05 10 2014 &   3.203 &    3.172 &    18.01 &    86.22 &   575.2 \\
  12.05 10 2014 &   3.197 &    3.181 &    18.00 &    86.25 &   576.7 \\  
  18.05 10 2014 &   3.158 &    3.231 &    17.90 &    86.40 &   585.8 \\  
  19.07 10 2014 &   3.152 &    3.239 &    17.87 &    86.41 &   587.3 \\  
  23.02 10 2014 &   3.126 &    3.270 &    17.71 &    86.45 &   593.0 \\  
  24.02 10 2014 &   3.120 &    3.278 &    17.66 &    86.46 &   594.4 \\
  25.07 10 2014 &   3.113 &    3.286 &    17.61 &    86.46 &   595.8 \\  
  26.01 10 2014 &   3.107 &    3.293 &    17.56 &    86.46 &   597.1 \\  
  15.05 11 2014 &   2.974 &    3.423 &    15.88 &    86.27 &   620.7 \\  
  17.05 11 2014 &   2.961 &    3.434 &    15.65 &    86.24 &   622.6 \\  
  18.02 11 2014 &   2.954 &    3.438 &    15.54 &    86.22 &   623.4 \\  
  19.02 11 2014 &   2.947 &    3.443 &    15.42 &    86.21 &   624.3 \\  
  20.06 11 2014 &   2.940 &    3.448 &    15.30 &    86.19 &   625.2 \\
  21.02 11 2014 &   2.934 &    3.453 &    15.18 &    86.18 &   626.0 \\  
  23.02 11 2014 &   2.920 &    3.461 &    14.93 &    86.15 &   627.6 \\  
  24.05 11 2014 &   2.913 &    3.466 &    14.79 &    86.14 &   628.4 \\
\hline                                  
\end{tabular}
\end{table*}

\end{appendix}

\end{document}